\documentclass[aps,pre,preprintnumbers,amsmath,twocolumn,amssymb,showpacs,longbibliography]{revtex4}
\usepackage{graphics}
\usepackage{bm}
\usepackage{dcolumn}
\usepackage{epsfig}
\usepackage{subfigure}
\usepackage{lipsum}
\usepackage{amsmath}
\usepackage{mathtools}
\usepackage{xcolor}
\usepackage{color}
\usepackage{bm}
\usepackage{amsmath,amssymb}
\usepackage{enumitem,kantlipsum}
\usepackage{lipsum}

\usepackage{gensymb}
\usepackage{graphicx}
\usepackage{psfrag}
\usepackage{color}
\usepackage{dcolumn}
\usepackage{bm}
\usepackage[normalem]{ulem}
\usepackage[absolute]{textpos}
\usepackage[normalem]{ulem}

\usepackage[utf8]{inputenc}

\def\noi{\noindent}
\def\bc{\begin{center}}
\def\ec{\end{center}}
 \newcommand{\bea}{\begin{equation}}
 \newcommand{\eea}{\end{equation}\noi}
 \newcommand{\ber}{\begin{eqnarray}}
 \newcommand{\eer}{\end{eqnarray}\noi}
\begin{document}
\title{Kardar-Parisi-Zhang universality in two-component driven diffusive models:
Symmetry and renormalization group perspectives}
\author{Pritha Dolai}\email{pritha.dolai@fau.de}
\affiliation{Friedrich-Alexander-Universit\"at Erlangen-N\"urnberg, 91054 Germany, 
Max-Planck-Zentrum für Physik und Medizin, Erlangen 91058, Germany}
 \author{Aditi Simha\footnote{Deceased}}
 \affiliation{Department of Physics, Indian Institute of Technology Madras, 
Chennai 600036, India}
\author{Abhik Basu}\email{abhik.123@gmail.com,abhik.basu@saha.ac.in} 
\affiliation{Theory Division, Saha Institute of Nuclear Physics, 1/AF 
Bidhannagar, Calcutta  700064, West Bengal, India}

\begin{abstract}
We elucidate the universal spatio-temporal scaling properties of the time-dependent 
correlation functions in a class of two-component one-dimensional (1D) driven diffusive 
system that consists of two coupled asymmetric exclusion processes. By using a 
perturbative renormalization group framework, we show that the relevant scaling 
exponents have values same as those for the 1D Kardar-Parisi-Zhang (KPZ) 
equation. We connect these universal scaling exponents with the symmetries of the model equations. We thus establish that these models belong to the 1D KPZ universality class.
\end{abstract}
\maketitle
\date{\today}

\section{Introduction}
Driven diffusive models are paradigmatic nonequilibrium models that display 
nonequilibrium universal scaling behavior different from any known dynamical 
scaling universality in equilibrium systems. One of the most well-known 
examples is the one-dimensional (1D) driven diffusive model~\cite{ddlg} that 
displays the Kardar-Parisi-Zhang (KPZ) universality class~\cite{kpz}. Subsequently, 
this has been generalized to a variety of multi-component driven diffusive 
models that yield a widely varying scaling behavior, ranging from continuously 
varying universality~\cite{abfreypre,abfreyjstat}, Kolmogorov scaling~\cite{abmhd} and weak dynamic scaling~\cite{mustansir} to 
strong dynamic scaling belonging to the KPZ universality class. In a recent 
study on the dynamical response to small distortions of a 1D
lattice drifting through a dissipative
medium about its uniform state, we show that the fluctuations, both transverse 
and longitudinal
to the direction of the drift, exhibit strong dynamic scaling and belong to the 
KPZ universality class ~\cite{ep_pre}. In spite of this large body of studies, the general 
question of universality in 1D multi-variable driven diffusive systems remains 
open. 

A particularly interesting class of models includes multi-species driven diffusive models that 
admit more than one conservation law. While there have been several studies in this context, a general consensus on the question of universality is still lacking. For instance, Ferrari et 
al.~\cite{ferrari}  studied a
two-species exclusion process by using a mode-coupling theory
and Monte Carlo simulations, and found two KPZ-modes. For a similar model, exact
finite-size scaling analysis of the spectrum indicates a
dynamical exponent $z = 3/2$~\cite{arita}, which is consistent with the 1D KPZ universality class. In earlier works on other
lattice gas models with two conservation laws, the presence
of a KPZ mode and a diffusive mode was claimed~\cite{mustansir,rakos}, 
indicating weak dynamic scaling. In Ref.~\cite{mustansir}, this occurrence of 
weak dynamic scaling is shown to be connected to special symmetries for carefully 
chosen parameters and the associated kinematic waves; else only KPZ 
universality is observed. This opens a question on how robust or general is 
the KPZ universality in 1D driven diffusive systems. Furthermore, this opens up to a broader generic issue of the robustness of a universality class against non-existence of a nonlinear coefficient - when vanishing of a particular nonlinear coefficient can affect the universality class obtained with a non-zero value of it and when cannot, and how this is connected to the symmetry of the model. A classic example is the $\phi^4$ Landau-Ginzburg theory with a $u\phi^4,\,u>0$, anharmonic term that describes the Ising model near its second order transition~\cite{chaikin}. However, if  $u=0$ identically, then the model becomes the Gaussian model, having critical exponents entirely different from the Ising model in the paramagnetic phase near the critical point, and in fact shows {\em no} phase transition~\cite{chaikin}. This opens the question whether a reverse scenario, where vanishing of a {\em relevant} (in the renormalization group or RG sense)  nonlinear coefficient in a dynamical model can leave the universal scaling and the universality class {\em unchanged}. In the absence of any general framework to study this in nonequilibrium systems, it is useful to construct simple models and perform explicit calculations to investigate this issue. There have already been some studies in this context by using various versions of driven diffusive generalized Burgers models. For instance, Ref.~\cite{popkov} proposed and studied a two-species driven diffusive model and found a KPZ mode and a non-KPZ mode with dynamic exponent $z=5/3$ within a mode coupling theory and dynamic Monte Carlo simulations. Subsequently, Ref.~\cite{spohn1} studied generalized coupled Burgers model and found the existence of a mode with $z=5/3$ in certain limits when some nonlinear coupling constants vanish by using mode coupling methods; see also Ref.~\cite{spohn2} in this context. More recently, Ref.~\cite{spohn3} revisited this generic class of coupled models and obtained within the framework of mode coupling theories one mode with $z=5/3$ in similar limits as the other previous related studies. Very recently, Schmidt {\em et al}~\cite{van} in a 1D three species model found two KPZ modes with $z=3/2$ and a third mode with $z=5/3$ by employing mode coupling methods and dynamical Monte Carlo simulations. The question that naturally arises is how or whether we can reconcile these results with the symmetries of the models and renormalization group (RG) perspectives.

In this work, we revisit this issue. We analyze a series of related two-species driven diffusive models~\cite{popkov,spohn1,ferrari,van}, and investigate the universal scaling properties of these models and their symmetries.
We analyze the model in Ref.~\cite{popkov}  using perturbative dynamic 
RG methods which is well-suited to extract universal scaling properties of dynamical models. We also  extend this analysis to the models in Refs.~\cite{spohn1,ferrari,van}.

Our RG treatment emphatically shows that, contrary to recent claims ~\cite{popkov}, there are only two KPZ-like modes admitted by 
the model discussed in Ref.~\cite{popkov}, implying 1D KPZ universal behavior. 

The rest of the article is organized 
as follows. In Sec.~\ref{model}, we describe the model. In Sec.~\ref{kpz}, we briefly discuss the KPZ equation and its universal properties in 1D. In Sec.~\ref{cont-2sp}, we set up the continuum equations that we study. We analyze the scaling properties of one particular case of the two-species model in Sec.~\ref{caseC}. In Sec.~\ref{rg}, we present the RG analysis of the model. In Sec.~\ref{summ}, we summarize. Some of the technical details are made available in Appendix for the interested reader.

\section{Multi-species models}\label{model}

We use the model studied in Ref.~\cite{popkov}, which is a stochastic lattice gas model of two 1D lattices or ``lanes'' with periodic boundary conditions, where particles can hop 
randomly on two lanes. Each lane has $N$ sites.
The model is defined as follows.
Particles move unidirectionally but without exchanging the lane. 
Periodic boundary conditions are imposed on each lane. Particles can hop from site $k$ 
to $k+1$ if site $k+1$ is empty. Furthermore, $n_{k}^{(i)}$ is the particle occupation number on site $k$ in lane $i$ and the particle hopping rate $r_{i}$ in lane $i$ from site $k$ to $k+1$ depends on the occupation numbers on sites $k$ and $k+1$ in the adjacent lane. The hopping rates in the two lanes are given by
\begin{equation}
 r_{1}=1+\gamma n^{(2)}/2,\,\,\,\, r_{2}=\tilde b+\gamma n^{(1)}/2\,,
\end{equation}
where the coupling parameter $\gamma \geq -min(1,\tilde b)$ and $n^{i}=n_k^{i} +n_{k+1}^i$ ~\cite{popkov}\,.
This model reduces to the two lane 
model of ~\cite{PopkovJstat} for $\tilde b=1$. 
Total number of particles $M_{i}$ in each lane is conserved. 
Thus, there are two conservation laws in this model.

\section{Kardar-Parisi-Zhang equation}\label{kpz}
We briefly revisit the KPZ equation~\cite{kpz} before embarking on our calculations.
Consider a 1D model with periodic boundary condition having one type of particles only that execute asymmetric exclusion processes. 
In a coarse-grained description, the local density $\rho(x,t)$ is known to follow the Burgers equation~\cite{fns}, which in turn 
can be mapped onto the KPZ equation for a single-valued height field $h(x,t)$
\begin{equation}
 \frac{\partial h}{\partial t} -\frac{\lambda}{2}(\partial_x h)^2 = \nu\frac{\partial^2 h}{\partial x^2} + \eta,\label{kpzeq}
\end{equation}
where $\rho(x,t) = -\partial_x h(x,t)$, and 
$\eta(x,t)$ is a zero-mean, Gaussian-distributed white noise. The correlator of $h$ shows 
universal spatio-temporal scaling in the long wavelength limit that are independent of the model parameters~\cite{barabasi}. In particular, 
in the Fourier space, correlator
\begin{equation}
 C_h(k,\omega)\sim \langle|h(k,\omega)|^2\rangle \sim q^{-1-2\chi_h}f_h(k^z/\omega).
\end{equation}
Here, $k$ and $\omega$, respectively, are Fourier wavevector and frequency; $\chi_h$, the roughness exponent and $z$, the dynamic exponent, 
are the universal scaling exponents, which characterize the universal scaling of $C_h(k,\omega)$; $f_h$ is a dimensionless scaling function. 
 For 1D KPZ, 
these exponents are known {\em exactly}: $\chi_h = 1/2$ and $z=3/2$~\cite{barabasi,natter}. 

{ Consider now a hypothetical dynamical equation
\begin{equation}
 \frac{\partial h}{\partial t} -\frac{\lambda_4}{2}(\partial_x h)^4 = \nu\frac{\partial^2 h}{\partial x^2} + \eta,\label{kpzeq4}
\end{equation}
where $\lambda_4$ is a coupling constant. Equation~(\ref{kpzeq4}) has the same symmetry as (\ref{kpzeq}). However, simple (but actually {\em incorrect}) power counting leads to the apparent  conclusion that $\lambda_4$ is irrelevant in the RG sense, which then should give the linear theory scaling to be the asymptotic long wavelength limit scaling in this model. However, this conclusion is {\em wrong}. One could write $(\partial_x h)^4=\langle (\partial_x h)^2\rangle (\partial_x h)^2$ to the leading order in fluctuations.  Evidently this generates a $(\partial_x h)^2$ in Eq.~(\ref{kpzeq4}), ultimately making it statistically identical to Eq.~(\ref{kpzeq}) in the long wavelength limit. That a $ (\partial_x h)^2$ is generated by fluctuations is not surprising, as (\ref{kpzeq}) and (\ref{kpzeq4}) are both nonlinear and belong to the same symmetry. Thus the lesson we can draw from this example is that the mere presence or absence of a particular relevant nonlinear term cannot be directly used to infer the universal properties; one needs to take the symmetries into account and start by using the most general equation containing all possible symmetry-permitted relevant nonlinear terms; see, e.g.,~\cite{foster}.}

\section{Continuum equations of motion for the two-species model}\label{cont-2sp}
 In order to extract the universal scaling behavior of the two-species model, we take the continuum hydrodynamic approach that is particularly suitable to extract long wavelength universal scaling properties~\cite{chaikin,foster}. In this approach, the equations of motion for the ``hydrodynamic variables'', which have diverging relaxation times in the vanishing wavevector limits,  are constructed in expansions in the fields and gradients, retaining {\em all possible} symmetry-permitted lowest order nonlinear terms. In the present study, the two conserved densities in the two lanes are the only hydrodynamic variables. 
We first consider the continuum coarse-grained equations of motion for the two densities in the two-species model. We begin by noting that
the stationary currents in two lanes are (see Ref.~\cite{popkov} for details) 
\begin{eqnarray}
  j_{1}(\rho_1,\rho_2)&=&\rho_1(1-\rho_1)(1+\gamma\rho_2)\,,\\
  j_{2}(\rho_1,\rho_2)&=&\rho_2(1-\rho_2)(b+\gamma\rho_1),
\end{eqnarray}
where $\rho_{i}=M_{i}/N$, $i=1,2$ are the particle densities in each lane ~\cite{popkov}\,.
Here $M_i$ is the number of particles on $i-$th lane and $N$ is the number of lattice sites in each lane.
The coarse-grained local density $\rho_i(x,t)$ for component $i$  obey the continuity equation which can be written in a compact form as 
\begin{equation}
\frac{\partial}{\partial t}{\rho_i}+A_{ij}\frac{\partial}{\partial x}{\rho_j}=0\,
\end{equation}
where $A$ is the Jacobian matrix with matrix elements $A_{ij}=\partial j_{i}/\partial \rho_{j}$. Local densities can be 
expanded around its stationary values: $\rho_{i}(x,t)=\rho_{i}+u_{i}(x,t)$\,. The normal modes are ${\phi_i}=R_{ij}{u_j}$ where $RAR^{-1}=\,$\text{diag}$(v_{i})$ with 
$v_{i}$'s being the eigenvalues of $A$ to the lowest order in spatial gradients; these $v_{i}$'s are the characteristic velocities with which local perturbations move. 
The transformation matrix $R$ satisfies the normalization condition $RCR^{T}$=1,  where $C$ is a  symmetric matrix; see Ref.~\cite{popkov} for a formal definition. 
Keeping the lowest order nonlinearities, the equations of motion of the normal modes are 
\begin{equation}
 \partial_t\phi_i=-\partial_x[v_i\phi_i+(1/2)\langle\vec{\phi},G^{(i)}\vec{\phi}\rangle-\partial_x(D\vec{\phi})_i+(B\vec{\eta})_i]
\end{equation}
with $i=1,2$\,. Here, $D$ and $B$ are the transformed diffusion matrix and transformed noise strength matrix respectively, and $\vec\eta$ is the noise vector.
 The mode coupling matrices, $G^{(i)}=(1/2)\sum_{j} R_{ij}(R^{-1})^{T}H^{(j)}R^{-1}$ depend on the Hessian matrix with elements 
$H^{(i)}=\partial^{2}j_{i}/(\partial \rho_{j}\partial \rho_{k})$\,. The coupling matrices $G^{(i)}$ are 
\[G^{(i)}=\left[\begin{array}{cc}
G_{11}^{(i)}   & G_{12}^{(i)}  \\
G_{21}^{(i)}   & G_{22}^{(i)}  \end{array} \right].\]  

In the equations of motion for $\phi_{1}$ and $\phi_{2}$ nonlinearities appear as an inner product $\langle\vec{\phi},G^{(i)}\vec{\phi}\rangle$ which can be expanded as 
\[\langle\vec{\phi},G^{(i)}\vec{\phi}\rangle= \left[ \begin{array}{cc} 
 \phi_{1}   & \phi_{2} 
 \end{array}\right]
 \left[\begin{array}{cc}
    G_{11}^{(i)}   & G_{12}^{(i)}   \\
    G_{21}^{(i)}   & G_{22}^{(i)}  \end{array}\right]   
\left[\begin{array}{c}
 \phi_{1} \\ \phi_{2} 
 \end{array}\right]  \]
$=G_{11}^{(i)}\phi_{1}^{2}+G_{12}^{(i)}\phi_{1}\phi_{2}+G_{21}^{(i)}\phi_{1}\phi_{2}+G_{22}^{(i)}\phi_{2}^{2}$\,.\\

We consider the normal modes for constant particle densities $\rho_{1}=\rho_{2}=\rho$ in each lane and coupling
parameter $\gamma=1$. The equations of motion for $\phi_{1}$ and $\phi_{2}$ are
\begin{equation}
\begin{aligned}
\partial_t \phi_{1} = {}  &-\partial_{x}(v_{1}\phi_{1})-\frac{1}{2}\partial_{x}[G_{11}^{(1)}\phi_{1}^{2}+(G_{12}^{(1)}+G_{21}^{(1)})\phi_{1}\phi_{2} \\
& +G_{22}^{(1)}\phi_{2}^{2}]+D_1\partial_{x}^{2}\phi_{1}-\partial_{x}(B_1\eta_{1}), \label{eq_phi1} 
 \end{aligned}
\end{equation}
\begin{equation}
\begin{aligned}
\partial_t \phi_{2} = {}  &-\partial_{x}(v_{2}\phi_{2})-\frac{1}{2}\partial_{x}[G_{11}^{(2)}\phi_{1}^{2}+(G_{12}^{(2)}+G_{21}^{(2)})\phi_{1}\phi_{2} \\
& +G_{22}^{(2)}\phi_{2}^{2}]+D_2\partial_{x}^{2}\phi_{2}-\partial_{x}(B_2\eta_{2}). \label{eq_phi2} 
 \end{aligned}
\end{equation}
We retain only the diagonal terms from the diffusion matrix since the pure diffusive terms have more dominant contributions than the cross diffusive terms to the eigenvalues (see Appendix for the details). 

The above equations are decoupled at the linear level. We consider the case for which $\tilde b=2$\,. In this case, the characteristic velocities are 
$v_1=1-\rho-3\rho^2$ and $v_2=2-3\rho-\rho^2$\,.  For $\tilde b=2$, $\gamma=1$, $\rho_1=\rho_2=\rho$ we can calculate the $G$ matrix using its definition above. 
The matrix elements of the coupling matrices are $G_{11}^{(1)}=-2g_0(6\rho^4-8\rho^3+5\rho^2+\rho-1)$\,, 
$G_{12}^{(1)}=G_{21}^{(1)}=g_0(4\rho^3-10\rho^2+8\rho-1)$\,, $G_{22}^{(1)}=-2g_0\rho(1-\rho)(2\rho^2-6\rho+3)$ 
and $G_{11}^{(2)}=4g_0\rho(1-\rho)$\,, $G_{12}^{(2)}=G_{21}^{(2)}=-g_0(1-2\rho^2)^2$\,, $G_{22}^{(2)}=4g_0(1-3\rho(1-\rho))$
with $g_0=-\frac{1}{2}\left[\rho(1-\rho)/(1-2\rho(1-\rho))^3\right]^{1/2}$ ~\cite{popkov}\,. 
We study three possible cases depending on the mode coupling matrix.\\
${\bf{A.}}$ If $G_{11}^{(1)}$ and $G_{22}^{(2)}$ are nonzero then the equations for $\phi_1$ and $\phi_2$ are same as the above equations (\ref{eq_phi1}) and (\ref{eq_phi2})\,. 
Now we make a change of variable $\phi_{1}=-\partial_{x} h_{1}$,  $\phi_{2}=-\partial_{x} h_{2}$ and rewrite the above equations  which will be of the form
\begin{equation}
\begin{aligned}
\frac{\partial h_{1}}{\partial t}={} & -v_{1}\frac{\partial h_{1}}{\partial x}+\frac{G_{11}^{(1)}}{2}(\frac{\partial h_{1}}{\partial x})^{2}+\frac{(G_{12}^{(1)}+G_{21}^{(1)})}{2}\frac{\partial h_{1}}{\partial x}\frac{\partial h_{2}}{\partial x}\\
  & +\frac{G_{22}^{(1)}}{2}(\frac{\partial h_{2}}{\partial x})^{2}+D_{1}\frac{\partial^{2}h_{1}}{\partial x^2}+B_{1}\eta_{1},\label{casea1}
 \end{aligned}
\end{equation}
\begin{equation}
\begin{aligned}
\frac{\partial h_{2}}{\partial t}={} & -v_{2}\frac{\partial h_{2}}{\partial x}+\frac{G_{11}^{(2)}}{2}(\frac{\partial h_{1}}{\partial x})^{2}+\frac{(G_{12}^{(2)}+G_{21}^{(2)})}{2}\frac{\partial h_{1}}{\partial x}\frac{\partial h_{2}}{\partial x}\\
  & +\frac{G_{22}^{(2)}}{2}(\frac{\partial h_{2}}{\partial x})^{2}+D_{2}\frac{\partial^{2}h_{2}}{\partial x^2}+B_{2}\eta_{2}.\label{caseb1}
 \end{aligned}
\end{equation}
Both of these equations contain nonlinear terms identical to those that appear in the 1D KPZ equation, in addition to other potentially relevant nonlinear terms. Equations~(\ref{casea1}) and (\ref{caseb1}) have {\em no particular symmetry}. 
These equations  were shown to exhibit KPZ dynamics in the long wavelength limit
~\cite{ep_pre}\,. Thus the dynamic exponents $z_{1}=z_{2}=3/2$, corresponding to {\em strong dynamic scaling}, together with the corresponding roughness exponent $\chi_1=\chi_2=1/2$, same as for the 1D KPZ equation. We note that both Eqs.~(\ref{casea1}) and (\ref{caseb1}) are invariant {\em only} under arbitrary constant shifts of $h_1$ and $h_2$, in addition to spatial translation and rotation; however, these equations have no invariance under inversions of space and/or the fields, i.e., no invariance under $x\rightarrow -x$ together or separately with $h_1\rightarrow -h_1,\,h_2\rightarrow -h_2$.\\

${\bf{B.}}$ If $G_{11}^{(1)}=G_{22}^{(1)}= 0$ but  $G_{22}^{(2)}\neq 0$, then with the same change of variables, these equations can be written as 
\begin{equation}
\begin{aligned}
 \partial_{t}h_{1}={} & -v_{1}\partial_{x} h_{1}+\frac{(G_{12}^{(1)}+G_{21}^{(1)})}{2}(\partial_{x} h_{1})(\partial_{x} h_{2})\\
 & +D_{1}\partial_{x}^{2}h_{1}+B_{1}\eta_{1},\label{h1caseb}
 \end{aligned}
\end{equation}
\begin{equation}
\begin{aligned}
 \partial_{t}h_{2}={} & -v_{2}\partial_{x} h_{2}+\frac{G_{11}^{(2)}}{2}(\partial_{x} h_{1})^{2}+\frac{G_{22}^{(2)}}{2}(\partial_{x} h_{2})^{2}\\
 & +\frac{(G_{12}^{(2)}+G_{21}^{(2)})}{2}(\partial_{x} h_{1})(\partial_{x} h_{2})+D_{2}\partial_{x}^{2}h_{2}+B_{2}\eta_{2}.\label{h2caseb}
 \end{aligned}
\end{equation}
Thus equation (\ref{h1caseb}) for $h_1$ does not contain any KPZ-like nonlinearity, but the corresponding equation (\ref{h2caseb}) still has a KPZ-like nonlinearity. { The overall symmetry of Eq.~(\ref{h1caseb}) and Eq.~(\ref{h2caseb}) are same as (\ref{casea1}) and (\ref{caseb1}) above. On symmetry ground, therefore, Eq.~(\ref{h1caseb}) and Eq.~(\ref{h2caseb}) should belong to the same universality class as (\ref{casea1}) and (\ref{caseb1}), which obviously has a larger set of model parameters. This expectation can be justified as follows.

We begin by noting that Eq.~(\ref{h1caseb}) and Eq.~(\ref{h2caseb}) have the same symmetry as the general equations (\ref{casea1}) and (\ref{caseb1}), which means no symmetry at all, except for the invariance under constant shifts of $h_1$ and $h_2$.   The equivalence between these two sets of equations may be established by using the general arguments given below. We note that Eq.~(\ref{h1caseb}) and Eq.~(\ref{h2caseb}) are obtained by expanding around the uniform states. This is justified so long as all the coefficients of the leading order terms are non-zero, as in that case all higher order nonlinear terms (i.e., nonlinear terms with more fields or more gradients) are irrelevant in the scaling/RG sense. If some of the coefficients of the leading order nonlinear terms vanish, then further careful analysis is required. Before we do that, we consider a simpler but well-known example. Consider the Landau-Ginzburg free energy for the Ising model near its critical point. It is is given by~\cite{chaikin}
 \begin{equation}
  {\cal F}_\text{Ising}= \int d^dx[\frac{1}{2}r\phi^2 +\frac{1}{2}({\boldsymbol\nabla}\phi)^2 + u\phi^4+v\phi^6].
 \end{equation}
 Here, $r=a(T-T_c)$, where $T_c$ is the mean-field critical temperature.
With $u>0$, the $v\phi^6$ term is irrelevant in the scaling/RG sense near the critical point. However, {\em if} we start with a bare scalar field free energy having $u=0$ but $v>0$, should {\em this} Ising model have a critical scaling behavior different from the conventional Ising universality class with $u>0$? While naively one is tempted to conclude that with $u=0$, the Ising spins should display universal scaling {\em different} from the standard Ising universality class, it is generally not true. It can be seen from the fact that we can write $\phi^6$ by decomposing it as $\langle \phi^2\rangle \phi^4$ (here $\langle...\rangle$ refers to an underlying equilibrium distribution), giving rise to an effective $\phi^4$-term in ${\cal F}_\text{Ising}$. While this seems contradictory to the expectation that the $\phi^6$-term is irrelevant (in the scaling/RG sense) near dimension $d=4$ for the Ising model, it is actually {\em not} contradictory. The coefficient of the fluctuation-generated $\phi^4$-term is finite near $d=4$, and hence would have no effect on the scaling {\em if} a $u\phi^4,\,u>0$-term had been present. In the absence of a $\phi^4$-term, i.e., with $u=0$, such a ``finite'' correction is actually ``infinitely'' larger than a corresponding bare coefficient $u$, as it vanishes. Hence, such a fluctuation-generated $\phi^4$-term cannot be discarded and instead should be retained.  This ultimately yields the standard Ising universality class for the second order transition.  This is not unexpected, since both $\phi^4$ and $\phi^6$ belong to the same symmetry, and even if the $\phi^4$-term is absent microscopically, it gets generated by the fluctuations from the $\phi^6$-term. Indeed, there is {\em no} symmetry ground to exclude the $\phi^4$-term, as its absence is {\em not} symmetry protected~\cite{tri}.
 Now coming back to our model, the absence of the term $(\partial h_2/\partial x)^2$-term is also {\em not} symmetry protected. Note that expanding around uniform steady states, one would naturally generate all possible symmetry-permitted nonlinear terms, including nonlinear terms which are higher than those kept in (\ref{casea1}) and (\ref{caseb1}) on the ground that they are irrelevant in the scaling/RG sense. However when some of the leading order nonlinear terms are absent (which is not symmetry protected), then further careful attention is necessary. Just as in our Ising model example discussed above, a lower order {\em relevant} nonlinear term can be generated from a higher order one in the present model. Such a fluctuation-generated lower order nonlinear term would be irrelevant if the theory originally included this lower order nonlinear term. However, if a particular lower order relevant nonlinear terms is {\em not} included, i.e., the corresponding nonlinear coupling constant vanishes, a corresponding fluctuation-generated contribution from the higher order nonlinear terms must be included for the same reasons as in our dicsussions on the Ising model with $u=0$. With this conceptual discussion in mind, we note that a possible such symmetry-permitted higher order  nonlinear term is $(\frac{\partial h_{1}}{\partial x})^2(\frac{\partial h_{2}}{\partial x})^2$ in the present model. We then replace $(\frac{\partial h_{1}}{\partial x})^2$ or $(\frac{\partial h_{2}}{\partial x})^2$ by their averages (over the steady states), thereby producing quadratic nonlinearities $(\frac{\partial h_{1}}{\partial x})^2$ or $(\frac{\partial h_{2}}{\partial x})^2$ in the respective equations of motion. Notice that if the corresponding lower order nonlinear terms were already present, such fluctuation-corrected contributions make no difference to the scaling.   Presenting the above argument in a different but equivalent way, we can phenomenologically assume  $v_2$ is not just a constant, but depends on $\partial h_2/\partial x$ via $v_2\sim v_2^0 + v_2^1\partial h_2/\partial x$, a possibility not ruled out by any symmetry considerations. This produces a $(\partial h_2/\partial x)^2$-term. We can also assume $U\equiv \frac{(G_{12}^{(2)}+G_{21}^{(2)})}{2}$ to depend on $\frac{\partial h_{1}}{\partial x}\frac{\partial h_{2}}{\partial x}$, which produces $(\frac{\partial h_{1}}{\partial x})^2(\frac{\partial h_{2}}{\partial x})^2$-term. As mentioned above, replacing $(\frac{\partial h_{1}}{\partial x})^2$ or $(\frac{\partial h_{2}}{\partial x})^2$ by their averages (over the steady states) produce quadratic nonlinearities $(\frac{\partial h_{1}}{\partial x})^2$ or $(\frac{\partial h_{2}}{\partial x})^2$ in the respective equations of motion. The resulting effective equations of motion are then identical in forms with the general equations (\ref{casea1}) and (\ref{caseb1}). We then conclude that Eqs.~(\ref{h1caseb}) and (\ref{h2caseb}) belong to the 1D KPZ universality class, similar to Eqs.~(\ref{casea1}) and (\ref{caseb1}).

Consider now a related case, if $G_{11}^{(1)}\neq 0$,  $G_{22}^{(2)}= 0$ and $G_{11}^{(2)}= 0$,  $G_{22}^{(1)}= 0$. The resulting dynamical equations are of the form 
\begin{equation}
\begin{aligned}
\frac{\partial h_{1}}{\partial t}={} & -v_{1}\frac{\partial h_{1}}{\partial x}+\frac{G_{11}^{(1)}}{2}(\frac{\partial h_{1}}{\partial x})^{2}+\frac{(G_{12}^{(1)}+G_{21}^{(1)})}{2}\frac{\partial h_{1}}{\partial x}\frac{\partial h_{2}}{\partial x}\\
  & +D_{1}\frac{\partial^{2}h_{1}}{\partial x^2}+B_{1}\eta_{1},\label{b1}
 \end{aligned}
\end{equation}
\begin{equation}
\begin{aligned}
\frac{\partial h_{2}}{\partial t}={} & -v_{2}\frac{\partial h_{2}}{\partial x}+\frac{(G_{12}^{(2)}+G_{21}^{(2)})}{2}\frac{\partial h_{1}}{\partial x}\frac{\partial h_{2}}{\partial x}\\
  & +D_{2}\frac{\partial^{2}h_{2}}{\partial x^2}+B_{2}\eta_{2}.\label{b2}
 \end{aligned}
\end{equation}
Since Eqs.~(\ref{b1}) and (\ref{b2}) have the same symmetry as the general equations (\ref{casea1}) and (\ref{caseb1}),
similar argument can be used to generate any missing terms in (\ref{b1}) and (\ref{b2}), relative to the general equations (\ref{casea1}) and (\ref{caseb1}).  These suggest that Eqs.~(\ref{b1}) and (\ref{b2}) should belong to the same universality class as (\ref{casea1}) and (\ref{caseb1}), a statement which can be justified by setting arguments similar to the previous example.

 There is yet another, more explicit way to demonstrate that Eqs.~(\ref{h1caseb}) and (\ref{h2caseb}) and also Eqs.~(\ref{b1}) and (\ref{b2}) belong to the same universality class as (\ref{casea1}) and (\ref{caseb1}). This is based on a perturbative approach that we illustrate with a specific case below.

}

A special case of (\ref{h1caseb}) and (\ref{h2caseb}) is when the cross-nonlinear term in (\ref{h2caseb}) vanishes, i.e., $G_{12}^{(2)}+G_{21}^{(2)}$ vanishes. In this case, equations (\ref{h1caseb}) and (\ref{h2caseb}) are jointly invariant under $x\rightarrow x,\,h_1\rightarrow -h_1,\,h_2\rightarrow h_2$, as has been  discussed in Das {\textit{et al}}~\cite{mustansir} in details. 
For this case, both dynamic RG and one-loop self-consistent (OLSC) schemes predict $z_{1}=2$ and $z_{2}=3/2$, implying {\em weak dynamic scaling}~\cite{mustansir}.\\

${\bf{C.}}$ If $G_{11}^{(1)}=0$ but $G_{22}^{(1)}\neq 0$ and $G_{22}^{(2)}\neq 0$ then the equation (\ref{eq_phi1}) for $\phi_{1}$ does 
not contain $\partial_x\phi_{1}^{2}$ term, but it contains the $\partial_x\phi_{2}^{2}$ term. We rewrite the equations  in terms of  $h_1$ and $h_2$ (as defined above) as
\begin{multline}
 \partial_{t}h_{1}+v_{1}\partial_{x} h_{1}+b_{1}(\partial_{x} h_{1})(\partial_{x} h_{2})\\
 +c_{1}(\partial_{x} h_{2})^{2}=D_{1}\partial_{x}^{2}h_{1}+B_{1}\eta_{1},\label{eq1}
\end{multline}
\begin{multline}
 \partial_{t}h_{2}+v_{2}\partial_{x} h_{2}+a_{2}(\partial_{x} h_{1})^{2}+b_{2}(\partial_{x} h_{1})(\partial_{x} h_{2})\\
 +c_{2}(\partial_{x} h_{2})^{2} =D_{2}\partial_{x}^{2}h_{2}+B_{2}\eta_{2},\label{eq2}
\end{multline}
where $b_{1}=-G_{12}^{(1)}$, $c_{1}=-\frac{G_{22}^{(1)}}{2}$, $a_{2}=-\frac{G_{11}^{(2)}}{2}$, $b_{2}=-G_{12}^{(2)}$ and 
$c_{2}=-\frac{G_{22}^{(2)}}{2}$. Notice that equations (\ref{eq1}) and (\ref{eq2}) have {\em no symmetry} except being invariant under constant shifts of $h_1$ and $h_2$, and have {\em two conservation laws} for the densities $\phi_1$ and $\phi_2$, which are {\em exactly same} as equations (\ref{casea1}) and (\ref{caseb1}). Principles of hydrodynamics,  which basically says that ``anything'' (i.e., any nonlinear term) that is symmetry-allowed will be generated, then tells us that the pair of equations (\ref{eq1}) and (\ref{eq2}) should belong to the same universality class as (\ref{casea1}) and (\ref{caseb1}), which is the 1D KPZ universality class for both $h_1$ and $h_2$. { Unexpectedly, in an OLSC study, }
$\phi_1$ is predicted to exhibit superdiffusive mode with 
dynamic exponent $z_{1}=5/3$ and $\phi_2$ shows 
KPZ dynamics with dynamic exponent $z_{2}=3/2$~\cite{popkov}; see also Refs.~\cite{spohn1,spohn2,spohn3} for related studies. 
 Clearly, Ref.~\cite{popkov} contradicts the general principles of hydrodynamics, which necessitates further detailed studies. Should equations (\ref{eq1}) and (\ref{eq2}) belong to the same 1D KPZ universality class, or not, is the question that we seek to answer below.
 
 { As in case B, we can phenomenologically argue that equations (\ref{eq1}) and (\ref{eq2}) should belong to the 1D KPZ universality class. To do this, we consider the possibility that in general the coefficient $b_1$, instead of being a constant, can be a function of $(\partial_x h_1)(\partial_x h_2)$, giving a nonlinear term $(\partial_x h_1)^2(\partial_x h_2)^2$. Now replacing $(\partial_x h_2)^2$ by $\langle (\partial_x h_2)^2\rangle$ produce a term of the form $(\partial_x h_1)^2$. Once this term is included, (\ref{eq1}) becomes identical with (\ref{casea1}) suggesting that the equations (\ref{eq1}) and (\ref{eq2}) should belong to the same universality class as (\ref{casea1}) and (\ref{caseb1}), which is the 1D KPZ universality class.} There are other similar phenomenological arguments, which lead to the same conclusion of 1D KPZ universality class; see also below.
 
 Alternatively, this can also be addressed by explicit use of perturbative approaches, which we discuss below.

\section{Universal scaling in case C} \label{caseC}

We focus on the scaling behavior in case C and systematically study the equations of motion. 
First we consider its linear limit, for which the correlation 
functions can be calculated {\em exactly}. The linearized equations for the normal modes have the generic forms
\begin{equation}
 \partial_{t}h_{1}+v_{1}\partial_x h_{1}=D_{1}\partial_x^{2}h_{1}+B_{1}\eta_{1},\label{lineq1}
\end{equation}
\begin{equation}
 \partial_{t}h_{2}+v_{2}\partial_x h_{2}=D_{2}\partial_x^{2}h_{2}+B_{2}\eta_{2}. \label{lineq2}
\end{equation}
Both Eqs.~(\ref{lineq1}) and (\ref{lineq2}) have underdamped kinematic waves. Since these equations are mutually decoupled, these waves can be removed 
in {\em both} Eqs.~(\ref{lineq1}) and (\ref{lineq2}) by going to the respective comoving frames. In the respective comoving frames, the 
correlation functions of $h_1$ and $h_2$ in the Fourier space then read
\begin{eqnarray}
 C_{h_1}({\bf q},\omega)=\langle |h_1 ({\bf q},\omega)|^2\rangle \sim q^{-1-2\chi_{1}}f_1(q^{z_1}/\omega),\\
 C_{h_2}({\bf q},\omega)=\langle |h_2 ({\bf q},\omega)|^2\rangle \sim q^{-1-2\chi_{2}}f_2(q^{z_2}/\omega)
\end{eqnarray}
in any dimension $d$. Unsurprisingly, these yield $\chi_1=1/2=\chi_2$  as 
the two 
roughness exponents in 1D, and $z_1=2=z_2$ as the two dynamic exponents in all $d$. We now set out to find how the nonlinear terms modify these exactly 
known values of the scaling exponents.

\subsection{Nonlinear effects on scaling in Case C}\label{rg}

Nonlinear terms, if relevant (in a scaling sense), alter the scaling exhibited by the linear theory. For instance, in 1D KPZ equation, 
dynamic exponent $z=2$ in the linear theory, whereas for the full nonlinear KPZ equation, $z=3/2$. Equations~(\ref{h1eq}) and (\ref{h2eq}) 
contain nonlinear terms that have structures similar to the nonlinear term in the KPZ equation. It is, thus, reasonable to expect that these 
nonlinear terms should change the scaling obtained in the linearized limit.

Before we embark upon any detailed analysis, we notice that due 
to the mutual couplings, there is {\em no} single frame where the kinematic wave terms in both Eqs.~(\ref{lineq1}) and (\ref{lineq2}) can be removed. 
We further note that Eqs.~(\ref{eq1}) and (\ref{eq2}) are invariant separately under the shift $h_1\rightarrow h_1 + const.,\,h_2\rightarrow h_2 + const.$.

Nonlinear terms preclude any exact analysis. Thus perturbative approaches are 
adopted. Naive perturvative approaches yield corrections to the model 
parameters that diverge in the long wavelength limit that can be 
successfully within the framework of the  dynamic RG
method~\cite{drg,Janssen}. We restrict here ourselves to low-order (one-loop) RG 
calculations. The perturbative expansion in powers of 
nonlinear 
coefficients results in diverging corrections in the long wavelength limit. 
These long wavelength divergences can be systematically treated by using a perturbative 
one-loop Wilson momentum shell dynamic RG~\cite{barabasi,Halperin, Halpin}\,. Dynamical 
fields $h_{1,2}({\bf{ q}},\omega)$ with higher wavevectors
($\Lambda/b<q<\Lambda,\,b>1$) 
are integrated out pertubatively up to the one-loop order, where $\Lambda$ is the upper wavevector cut off. Wavevectors are rescaled as $q'=bq$ 
so that the upper wavevector cut off is restored to $\Lambda$\,.  We perform the 
following scale transformations:
$x\rightarrow bx$ (corresponding to $q'=bq$), $t\rightarrow b^{z}t$, $h_{1}\rightarrow b^{\chi_{1}}  h_{1}$ 
and $h_{2}\rightarrow b^{\chi_{2}}h_{2}$\,. 

\begin{figure}[!h]
\begin{center}
{ \includegraphics[scale=0.25,keepaspectratio=true]{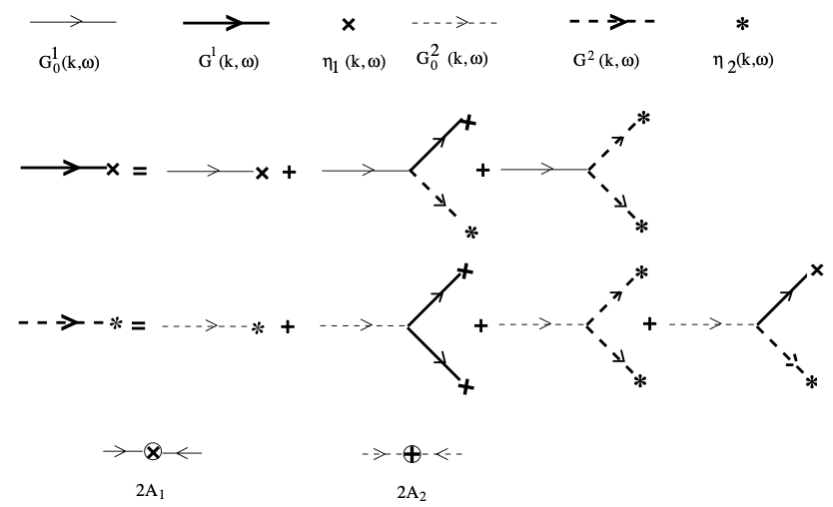}
}
\caption{Symbols that are used in the RG calculations. Solid lines correspond to $h_1$ field and dashed lines correspond to $h_2$ field.}
\label{symbol}
\end{center}
\end{figure}

In the comoving frame of $h_{1}$, the dynamical equations (\ref{eq1}) and (\ref{eq2}) take the form
\begin{equation}
 \partial_{t}h_{1}+b_{1}(\partial_x h_{1})(\partial_x h_{2})+ c_{1}(\partial_x 
h_{2})^{2}=D_{1}\partial_x^{2}h_{1}+B_{1}\eta_{1}\,\label{h1eq}
\end{equation}
\begin{multline}
 \partial_{t}h_{2}+v\partial_x h_{2}+a_{2}(\partial_x h_{1})^{2} +b_{2}(\partial_x h_{1})(\partial_x h_{2})\\
+c_{2}(\partial_x h_{2})^{2} =D_{2}\partial_x^{2}h_{2}+B_{2}\eta_{2}\,\label{h2eq}
\end{multline}
where the relative wave speed $v=v_2-v_1$\,. The noises $\eta_1$ and $\eta_2$ are assumed to be zero-mean Gaussian distributed with variances
\begin{eqnarray}
\langle\eta_{1}(x,t)\eta_{1}(x',t')\rangle&=&2A_{1}\delta(x-x')\delta(t-t'), \\
\langle\eta_{2}(x,t)\eta_{2}(x',t')\rangle&=&2A_{2}\delta(x-x')\delta(t-t')\,.
\end{eqnarray}
We need to solve Eqs.~(\ref{h1eq}-\ref{h2eq}) perturbatively and obtain the fluctuation corrections to the model parameters. Symbols that are used in dynamic RG calculations are shown in Fig.~\ref{symbol}\,.

\begin{figure}[!h]
\begin{center}
{ \includegraphics[scale=0.25,keepaspectratio=true]{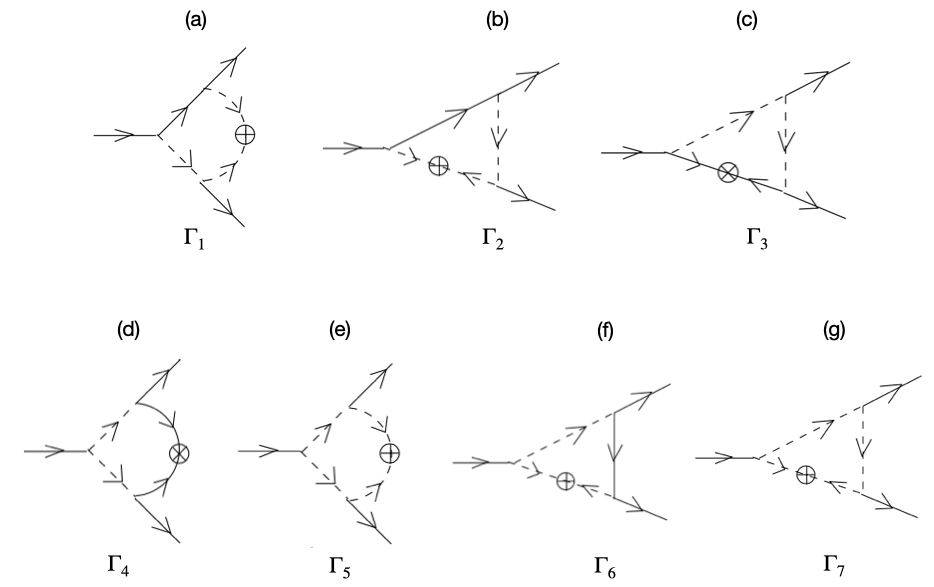}
}
\caption{Generation of KPZ-like nonlinear terms with coefficient $a_1$ in Eq.~\ref{h1eq1} in the iterative expansions of Eq.~\ref{h1eq}. 
Terms (a)-(c)  originate from the vertex $b_1$, terms (d)-(g) are generated from the vertex $c_1$\,.}
\label{KPZ-like}
\end{center}
\end{figure}

From the linear part of the equations (\ref{h1eq}-\ref{h2eq}) we calculate $G_{0}^{1}(k,\omega)$ and $G_{0}^{2}(k,\omega)$\,,  
the bare propagators of $h_{1}(k,\omega)$ and $h_{2}(k,\omega)$ respectively. 
Bare propagators are 
\begin{eqnarray}
G_{0}^{1}(k,\omega)=\frac{1}{D_{1}k^{2}+i\omega},\\
G_{0}^{2}(k,\omega)=\frac{1}{D_{2}k^{2}+i(\omega-kv)}.
\end{eqnarray}
Now notice that Eq.~(\ref{h1eq}) for the dynamics of $h_1$ does not contain the
KPZ-like nonlinear term $(\partial_x h_1)^2$, although it is {\em not} 
prohibited by any symmetry arguments. In contrast, Eq.~(\ref{h2eq}) contains 
all possible symmetry permitted bilinear nonlinearities of $\partial_x h_1$ and $\partial_x 
h_2$. Unsurprisingly, in an iterative, bare perturbation theory such 
KPZ-like nonlinear terms are indeed generated in Eq.~(\ref{h1eq}). These terms are represented graphically in 
Fig.~\ref{KPZ-like}\,. Let $a_1$ be the effective coefficient of the term $(\partial_x h_1)^2$ that is generated in iterative expansions. 
We are thus obliged to add this term to Eq.~(\ref{h1eq}) for consistency reasons {\em before} applications of RG methods:
\begin{multline}
 \partial_{t}h_{1}+a_1(\partial_x h_1)^2+b_{1}(\partial_x h_{1})(\partial_x h_{2})\\
 + c_{1}(\partial_x h_{2})^{2}=D_{1}\partial_x^{2}h_{1}+B_{1}\eta_{1}\,\label{h1eq1}
\end{multline}
The form of $a_1$ as obtained in the lowest order iterative expansion is  
given in Appendix. 
That a term of the form $(\partial_x h_1)^2$ is generated under iterative
expansion is not surprising. Equations~(\ref{h1eq}) and (\ref{h1eq1}) have the 
same symmetries; in other words the absence of $(\partial_x h_1)^2$ in 
(\ref{h1eq}) is {\em not symmetry-protected}, i.e., there is no symmetry
that forbids the existence of this term in (\ref{h1eq}). Since Eqs.~(\ref{h1eq1}) together with (\ref{h2eq}) and correspondingly the pair Eqs.~(\ref{h1eq}) and (\ref{h2eq}) have exactly the same symmetries, they must belong to the same universality class. Furthermore, from the explicit forms of the one-loop expressions as given in Appendix, it is clear that these are {\em inhomogeneous}, which means $a_1=0$ is not a fixed point at all of this model.  {\em This remains an important technical conclusion from the present study.}

The presence of the term $(\partial_x h_1)^2$ in Eq.~(\ref{h1eq1}) has been motivated phenomenologically above. Alternatively, it can be further motivated phenomenologically as follows. Consider Eq.~(\ref{eq1}). Since the hydrodynamic equations are actually written down by expanding around uniform steady states assuming small fluctuations, the phenomenological coefficient $v_1$, assumed constant here, can actually depend upon the local fields in ways that respect the overall symmetries of the dynamics, i.e., invariance under constant shifts of $h_1$ and $h_2$. This consideration allows us to generalize $v_1$ and replace it by  a field-dependent coefficient $v_1'\equiv v_1 + \tilde a_1 \partial_x h_1$, where $\tilde a_1$ is a phenomenological constant coefficient. We now use this in Eq.~(\ref{eq1}); the resulting equation has the same form as Eq.~(\ref{h1eq1}). Here, two short technical comments are in order: (i) The coefficient $v_1'$ can depend not only on $\partial_x h_1$, but also on $\partial_x h_2$ and all other terms that involve more fields and/or more derivatives. Inclusion of these terms do not generate any new relevant (in the RG sense) terms in the resulting final equations. (ii) Secondly, all the other coefficients in Eqs.~(\ref{eq1}) and (\ref{eq2}) too can depend on the local fields, subject to the overall symmetries. Again, inclusion of these contributions do not affect the physics in the long wavelength limit. We, therefore, ignore all these irrelevant contributions.

We work with the following {\em effective} equations, which are the most general equations with all possible symmetry-permitted nonlinear terms, written in the comoving frame of $h_1$:
\begin{multline}
  \partial_{t}h_{1}+a_1(\partial_x h_1)^2+b_1(\partial_xh_{1})(\partial_x h_{2})\\
 +c_1(\partial_x h_{2})^{2}=D_{1}\partial_x^{2}h_{1}+\eta_{1},\label{modeq1}
\end{multline}
\begin{multline}
 \partial_{t}h_{2}+v\partial_x h_{2}+a_2(\partial_x h_{1})^{2} +b_{2}(\partial_x h_{1})(\partial_x h_{2})\\
+c_{2}(\partial_x h_{2})^{2}=D_{2}\partial_x^{2}h_{2}+\eta_{2},\label{modeq2}
\end{multline}
where the factors of $B_1$ and $B_2$ have been absorbed in $\eta_1$ and $\eta_2$, respectively, for convenience, and $v\equiv v_2-v_1$. 
Since a term $(\partial_x h_1)^2$ is symmetry-permitted and is generated under iterative expansion, and has the same scaling dimension as the existing nonlinear terms~\cite{ep_pre}, any RG analysis must start with an equation that already includes this term, even though the naive hydrodynamic theory may not have it. Thus Eqs.~(\ref{modeq1}) and (\ref{modeq2}) should be the starting equations for any dynamic RG analysis for this problem. 
Equations (\ref{modeq1}) and (\ref{modeq2}) are identical to those studied in 
Ref.~\cite{ep_pre} by means of RG and Monte-Carlo simulations. We do not repeat the calculation which is straightforward and instead refer the reader to Ref.~\cite{ep_pre} for the details, and briefly revisit the conclusions of Ref.~\cite{ep_pre} below. 
It has been shown in Ref.~\cite{ep_pre} 
that the model belongs to the 1D KPZ universality class where 
$\chi_1=\chi_2=1/2$, together with a {\em single} dynamic exponent 
$z=3/2$. These led us to conclude that case C in the present study belongs to 
the 1D KPZ universality as well. 

Our scheme of calculations follow the scheme outlined in Refs.~\cite{mustansir,ep_pre}. Under 
perturbative RG, no new relevant terms are further generated. We obtain the fluctuation corrections up to the one-loop order. As in 
Ref.~\cite{ep_pre}, each model parameter receives corrections that either 
originate from only the KPZ-like nonlinearities result (equivalently, from one-loop diagrams that are identical to those in the RG for the KPZ equation), 
or nonlinearities other than the KPZ-type. For reasons identical to those elaborated in Ref.~\cite{ep_pre}, the former class of the diagrams are {\em more} 
relevant than the other diagrams in the long wavelength limit. As in Ref.~\cite{ep_pre} coupling constants do not receive any fluctuation corrections at the 
one-loop order in the long wavelength limit. We retain only these most dominant 
corrections.  Rescaling space, time and fields as mentioned above the 
different RG flow equations read
\begin{eqnarray}
\frac{dD_{1}}{dl} &=& D_{1}[z-2+g]\,, \nonumber \\
\frac{dA_{1}}{dl} &=& A_{1}[z-1-2\chi_{1}+g]\,, \nonumber \\
\frac{dD_{2}}{dl} &=& D_{2}[z-2+\frac{1}{2}mnrg]\,, \nonumber \\
\frac{dA_{2}}{dl} &=& A_{2}[z-1-2\chi_{2}+pn^{2}g]\,, 
\end{eqnarray}
where the coupling constant $g= \frac{A_{1}a_{1}^{2}}{\pi D_{1}^{3}}$ and 
dimensionless constants $m=\frac{D_{1}}{D_{2}}$, 
 $p=\frac{A_{1}}{A_{2}}$, $n=\frac{a_{2}}{a_{1}}$, and $r=\frac{b_{1}}{a_{1}}$\,. Unsurprisingly, these are identical to those derived in Ref.~\cite{ep_pre}. We therefore conclude $z=3/2,\,\chi_1=\chi_2=1/2$. Thus, the model belongs to the KPZ universality class.

It remains to be seen how one may arrive at the same conclusion 
from an OLSC study of case C. In an OLSC treatment, in principle
one is 
required to solve {\em all} the  correlation functions and propagators, and also the nonlinear vertices 
{\em self-consistently}, which  receive corrections that are unbounded or diverge 
in the thermodynamic limit at the one-loop order relative to their {\em bare} values in the 
theory~\cite{amit-ab-jkb}. OLSC further necessitates that at the one-loop order 
no {\em new} term should appear that would change the OLSC scaling if it were already
present in the original theory. This consideration yields $z=3/2$ and 
$\chi=1/2$ for the 1D KPZ equation~\cite{mustansir}. Applying this to case C here, we 
note that Eq.~(\ref{h1eq}) does not contain any $(\partial_x h_1)^2$ term, i.e., 
the ``bare'' value of $a_1$ is zero. On the other hand, as stated above, a 
term of the form $(\partial_x h_1)^2$ with a finite coupling coefficient 
$a_1$ is generated at the one-loop order. This conclusion remains unchanged 
 even when one uses the self-consistent scaling forms for the correlation 
functions and the propagators. Since the ``bare'' value of $a_1$ is zero, 
self-consistent generation of a finite $a_1$ at the one-loop order implies 
a generation of a one-loop correction that, relative to its bare value, is 
infinitely large, and hence cannot be dropped. In other words, {\em not} 
retaining this correction would result a {\em non self-consistent} solution. 
Since the presence of a ``bare'' term of the form $(\partial_x h_1)^2$ 
can affect the scaling exponents, we are required to include it before 
embarking on OLSC calculations. Once included, the OLSC calculations should reproduce 
~\cite{ep_pre} directly that give $z=3/2,\chi_1=\chi_2=1/2$, in agreement with the 
RG analysis discussed above.

Finally, we make one technical comment. We have argued above that even if the ``bare'' value of $a_1$ is zero, it is generated due to fluctuation effects. However, in order for it to be effective or relevant, it must be large enough, i.e., ${\cal O}(1)$. The size of the fluctuation-induced $a_1$, as shown in Appendix, depends upon all the other parameters and also on the system size. We thus expect that to extract the true asymptotic long wavelength scaling behavior, the system size should be sufficiently large enough. Further numerical investigations should be helpful in this context. 

\section{Summary and outlook}\label{summ}

In summary, we have revisited the universal scaling properties of the density fluctuations in  two-species periodic asymmetric exclusion processes. We argue that the continuum hydrodynamic equations of motion that one naively writes down may not include all the symmetry-permitted nonlinear terms. In fact, simple perturbative expansions produce any missing nonlinearity, so long it is symmetry-permitted, that is as relevant as the existing nonlinear terms. Hence, we argue that it should be included before undertaking any dynamic RG analysis on this model. Once this nonlinearity is considered, the effective equations of motion become identical to those analyzed in Ref.~\cite{ep_pre}. This in fact immediately allows us to conclude that the universal scaling exponents are identical to those for the 1D KPZ equation. That we get 1D KPZ scaling, in spite of having two conservation laws (corresponding to two conserved species) should not be surprising. This happens because, the coupled equations that include the extra nonlinear term, added on symmetry grounds, decouple in the long wavelength limit into two independent 1D KPZ equations ~\cite{mustansir}. 
We note in the passing that our conclusions are at variance with Ref.~\cite{popkov}, where  OLSC method was employed to calculate the universal scaling exponents. They did not consider the missing symmetry-permitted nonlinear term while applying OLSC, and obtained different values of the scaling exponents. Indeed, if the OLSC method is employed {\em after} inclusion of this missing nonlinear term, it should yield the same results as here, for reasons similar to Ref.~\cite{mustansir}.  Our results can be confirmed by solving the stochastically driven hydrodynamic equations numerically. Lastly, it is possible to explore scaling in the renormalized theory, after dropping renormalized versions of some of the coefficients,  e.g., by imposing renormalized  $G_{11}^{(1)}$ or
$G_{22}^{(1)}$ to zero. These would correspond to special points in the parameter space and would be accessible only by additional fine tuning, akin to accessing a tricritical point in equilibrium systems. The long wavelength scaling behavior at these special points can in general be different. 

The OLSC method used elsewhere intrinsically ignores the vertex corrections, and can work well where the vertex correction is absent due to some symmetry reasons, e.g., the KPZ equation. In the present problem, there are no symmetry reasons that prohibit fluctuation corrections to all the coupling constants. However, due to the presence of the underdamped waves, the vertex corrections turn out to be {\em finite}~\cite{ep_pre}, and hence can be ignored so far as the universal scaling is concerned. The RG framework can handle the vertex corrections, whether relevant (in the RG sense), or not - systematically. The generic question to what degree the universal scaling is affected by a missing coupling constant can occur in a theory with relevant vertex corrections as well. In such cases,  applications of RG methods would in fact be unavoidable. 

Our work can be extended in a variety of ways. It will be interesting to extend our analysis to multi component models; see, e.g., Ref.~\cite{van}. It will also be interesting to construct a suitable higher dimensional version of the two-species asymmetric exclusion and obtain the corresponding higher dimensional hydrodynamic equations. One can then ask whether the higher dimensional version of the present model belongs to the higher dimensional KPZ universality class. Secondly, one can introduce ``mass conserving reactions'' of various kinds, that will reduce the number of conservation laws from two to one in the present model. How that affects the long wavelength universal scaling properties is an interesting question to study in the future. One may additionally introduce lane exchanges by the particles and see whether the conclusions drawn here still remain valid in the presence of exchange. Effects of quenched disorder on the universal scaling properties can also be studied. It has recently been shown that in a single-component periodic TASEP with quenched disordered hopping rates, the disorder is irrelevant (in a RG sense) when the system is away from half-filling, and the scaling properties of the fluctuations in the long wavelength limit belong to the 1D KPZ universality class. In contrast, close to half-filling a new universality class emerges~\cite{astik-prr}. We hope our work will inspire future theoretical work along these directions.


\appendix

\section{Linear stability analysis: Contribution of the pure diffusive terms}
We consider Eqs.~(\ref{eq1}-\ref{eq2}) with cross-diffusive terms. These equations can be written as 

 \begin{multline}
 \partial_{t}h_{1}+v_{1}\partial_{x} h_{1}+b_{1}(\partial_{x} h_{1})(\partial_{x} h_{2})\\
 +c_{1}(\partial_{x} h_{2})^{2}=D_{11}\partial_{x}^{2}h_{1}+D_{12}\partial_{x}^{2}h_{2} +B_{1}\eta_{1},
\end{multline}
\begin{multline}
 \partial_{t}h_{2}+v_{2}\partial_{x} h_{2}+a_{2}(\partial_{x} h_{1})^{2}+b_{2}(\partial_{x} h_{1})(\partial_{x} h_{2})\\
 +c_{2}(\partial_{x} h_{2})^{2}=D_{21}\partial_{x}^{2}h_{1} +D_{22}\partial_{x}^{2}h_{2}+B_{2}\eta_{2}\,.
\end{multline}
Fourier transforming the both sides of the above equations we get
\[ \left[ \begin{array}{c} 
 \partial_t\tilde{h}_{1}   \\ \partial_t\tilde{h}_{2} 
 \end{array}\right]=  M^{-1}
\left[\begin{array}{c}
 B_{1}\eta_1 \\ B_2\eta_2 
 \end{array}\right],  \]
where $\tilde{h}_1$, $\tilde{h}_2$ are the Fourier transforms of $h_1$ and $h_2$ respectively and 
\[ M= \left[\begin{array}{cc}
 -ikv_1 +D_{11}k^2  & D_{12}k^2 \\
D_{21}k^2 &  -ikv_2+D_{22}k^2 \\
\end{array}\right]. \]
Eigenvalues of matrix $M$ are 
\begin{multline}
\lambda_{1,2}= \frac{1}{2} \left[   -ik(v_1 +v_2) + (D_{11}+D_{22})k^2 \right]\\
\pm \frac{1}{2}\sqrt{\alpha_1 k^2 +2i\alpha_2 k^3 +\alpha_4 k^4}
\end{multline}
$\alpha_1=-(v_1-v_2)^2$, $\alpha_2=(v_1D_{22}+v_2D_{11}-v_1D_{11}-v_2D_{22})$\,, $\alpha_4=(-4D_{12}D_{21}+(D_{11}-D_{22})^2)$\,.
For small $k$ , terms containing $k^4$ and $k^3$ are negligible compared to that with $k$. For small $k$, the eigenvalues are
\begin{equation}
\lambda_1=\frac{1}{2}\left[ -2ikv_2 +(D_{11}+D_{22})k^2 \right]
\end{equation}
\begin{equation}
\lambda_2=\frac{1}{2}\left[-2ikv_1 +(D_{11}+D_{22})k^2 \right]
\end{equation}
This confirms that the pure diffusive terms have more dominant contributions than the cross diffusive terms to the eigenvalues.
\section{Generation of the  KPZ-like nonlinear term with coefficient $a_1$ at one loop order }

We give below the value of the coefficient $a_1$ in Eq.~(\ref{h1eq1}) obtained in the one loop expansions of Eq.~(\ref{h1eq}); see Fig.~\ref{KPZ-like}.
\begin{equation}
 a_1=\Gamma_1+\Gamma_2+\Gamma_3+\Gamma_4+\Gamma_5+\Gamma_6+\Gamma_7,
\end{equation}
where
\begin{widetext}
\begin{eqnarray}
\hspace{-2.0cm}
\Gamma_1=\frac{2A_2b_1b_2}{\pi D_2^2} &\int\frac{(4q^2-k_1^2)dq}{\left[k_1^2+4k_2^2-4k_1q-8k_2q+8q^2\right]}\\ \nonumber
&\times\frac{1}{\left[ 4D_2(k_2-q)^2-2iv(k_1+2q)+D_1(k_1+2q)^2\right]}\\ \nonumber
& \times \frac{\left[-2iv(k_1+2q)+D_1(k_1+2q)^2+D_2(k_1^2+8k_2^2-4k_1q-16k_2q+12q^2)\right]}{\left[D_2(k_1-2q)^2+D_1(k_1+2q)^2-2iv(k_1+2q)\right]}
\end{eqnarray}
\begin{eqnarray}
\hspace{-2.2cm}
\Gamma_2=&-\frac{4A_2b_1b_2}{\pi D_2^2}\int\frac{(k_2+q)(k_1+2q)dq}{\left[k_1^2+4k_2^2-4k_1q+8q(k_2+q)\right]}\\
&\times \frac{1}{\left[D_2(k_1-2q)^2-2iv(k_1+2q)+D_1(k_1+2q)^2\right] }
\end{eqnarray}
\begin{eqnarray}
\hspace{-2.2cm}
 \Gamma_3=&\frac{16A_1a_2b_2}{\pi} \int\frac{(k_1-2q)^2(k_2+q)(k_1+2q)}{\left[2v(k_1-2q)-i(D_1(k_1-2q)^2-4D_2(k_2+q)^2)\right]} \\ \nonumber
& \times \frac{ \left[-4iv(k_1-2q)+D_2(k_1^2+4k_2^2+4k_1q+8q(k_2+q)\right]}{\left[2v(k_1-2q)+i(D_1(k_1-2q)^2+4D_2(k_2+q)^2)\right]} \\ \nonumber
& \times \frac{1}{\left[2v(k_1-2q)-i\left[D_1(k_1-2q)^2-D_2(k_1+2q)^2\right]\right]}\\ \nonumber
& \times \frac{1}{\left[2v(k_1-2q)+i\left[D_1(k_1-2q)^2+D_2(k_1+2q)^2\right]\right]}
\end{eqnarray}
\begin{eqnarray}
\hspace{-2.2cm}
\Gamma_4=&\frac{32A_1a_2^2}{\pi}\int \frac{(k_1^2-4q^2)(q-k_2)^2dq}{D_2(k_1^2+4q^2)\left[4(-iv+D_1(k_2-q))(k_2-q)+D_2(k_1+2q)^2\right]}\\ \nonumber
& \times \frac{1}{\left[-4(iv+D_1(k_2-q))(k_2-q)+D_2(k_1+2q)^2\right]}
\end{eqnarray}
\begin{eqnarray}
\hspace{-2.2cm}
\Gamma_5=&\frac{2A_2b_2^2}{\pi}\int \frac{(4q^2-k_1^2)dq}{D_2^3(k_1^2+4q^2)\left[k_1^2+4k_2^2-4(k_1+2k_2)q+8q^2\right]} \\ \nonumber 
& \times \frac{(k_1^2+4k_2^2-8k_2q+8q^2)}{\left[k_1^2+4k_2^2+4(k_1-2k_2)q+8q^2\right]}
\end{eqnarray}
\begin{equation}
\hspace{-2.2cm}
\Gamma_6=-\frac{8A_2a_2b_1}{\pi}\int\frac{(k_2+q)(k_1+2q)dq}{D_2^2(k_1^2+4q^2)\left[D_2(k_1-2q)^2+4(k_2+q)(-iv+D_1(k_2+q))\right]}
\end{equation}
\begin{equation}
\hspace{-2.2cm}
\Gamma_7=-\frac{4A_2b_2^2}{\pi}\int \frac{(k_2+q)(k_1+2q)dq}{D_2^3(k_1^2+4q^2)\left[k_1^2+4k_2^2-4k_1q+8q(k_2+q)\right]}
\end{equation}
\end{widetext}


\begin{thebibliography}{10}
\bibitem{tri} If one sets renormalized $u=0$, one would obtain the tricritical scaling at the tricritical point different from the usual Ising universality class. Accessing this requires additional parameter tuning. We are not concerned with such tricritical point-like behavior in the present model.
\end{thebibliography}

\end{document}